# Thermostatic Hyperthermia with Non-invasive Temperature Monitoring through Speed of Sound Imaging

Yuchan Wang, Yuening Wang, Yuan Jie and Paul Carson

*Abstract*—Hyperthermia therapy (HT) is used to treat diseases through heating of high temperature usually in conjunction with some other medical therapeutics like chemotherapy and radiotherapy. In this study, we propose a promising thermostatic hyperthermia method that uses high intensity focused ultrasound (HIFU) for clinical tumor treatment combined with diagnostic ultrasound image guidance and non-invasive temperature monitoring through speed of sound (SOS) imaging. HIFU heating is realized by a ring ultrasound transducer array with 256 elements. The inner structure information of thigh tissue is obtained by B-mode ultrasound imaging. Since the relationship between the temperature and the SOS in the different human tissue is available, the temperature detection is converted to the SOS detection obtained by the full-wave inversion (FWI) method. Simulation results show that our model can achieve expected thermostatic hyperthermia on tumor target with 0.2°C maximum temperature fluctuation for 5 hours. This study verifies the feasibility of the proposed thermostatic hyperthermia model. Furthermore, the temperature measurement can share the same ultrasound transducer array for HIFU heating and B-mode ultrasound imaging, which provides a guiding significance for clinical application.

*Index Terms*—HIFU, ring ultrasound transducer array, thermostatic hyperthermia

## I. Introduction

HYPERTHERMIA therapy (HT) is used to treat diseases through keeping heating the whole body or a part in high temperature for a long period, which is usually used alone or in conjunction with some other medical therapeutics like chemotherapy and radiotherapy [1]. Many successful clinical trials have applied the therapy to treat different types of cancer tumors, such as recurrent breast cancer, liver cancer, uterine fibroids and bladder cancer [2-5]. HT is mainly divided into three categories: local HT, regional HT and whole-body HT [6]. The purpose of local HT is to increase the temperature of the tumor without affecting the surrounding normal tissues. Also, local HT is generally applied to treat tumors below the skins or in body cavities near the surface of the body through using external or interstitial heating modalities [7]. In contrast, regional HT is more complicated than local HT because its targets of treatment are deep-seated tumors in body cavities, especially locally advanced tumors. The ways to achieve regional HT are non-invasive methods including heating the blood, irrigating the body cavities and radiofrequency (RF). Whole-body HT is usually applied to treat metastatic cancer by using surface heating, radiation induction or extracorporeal induction. During the entire hyperthermia treatment, the temperature of the tumor and surrounding tissues needs to be strictly controlled at the expected temperature. Among many techniques such as microwave, infrared and RF, high intensity focused ultrasound (HIFU) is increasingly being developed as a promising modality to induce hyperthermia for clinical tumor treatment combined with diagnostic ultrasound guidance [8-9].

For many kinds of benign and malignant tumors, HIFU plays an auxiliary role in the current standard treatments such as surgery, gene therapy and radiation. HIFU uses the characteristic that ultrasound can penetrate human body without damage and focus on vivo causing thermal effect (main effect), cavitation effect and mechanical effect. Since the high temperature can destroy the cell structure and damage the protein, the high temperature generated by thermal effect on the focus will lead to the coagulation necrosis of the tumor target [10-12] instead of burning out the surrounding normal tissue outside the focus. The sensor emits HIFU signals to the tissue target and then the tissue absorbs ultrasonic energy to convert it into thermal energy during tumor therapy. This heat deposition can cause the tissue temperature to rise rapidly. Due to the focusing of the ultrasound beam, thermal power is mainly added to the tumor target of a small area at the desired range of temperatures (e.g., 5°C-6°C above the ambient temperature of 37°C), which means there is almost no obvious heat deposition on the surrounding tissue so as to avoid the unpredictable growth of healthy tissues caused by excessive temperature [13].

There are two typical types of HIFU equipment systems that have been put into clinical practice for treatment of tumors: HIFU systems guided by magnetic resonance imaging (MRgHIFU) and HIFU systems guided by Ultrasound (USgHIFU). The high spatial resolution of MRgHIFU systems enables the accuracy of image guidance, which offers abundant anatomic details for tumor detection [14]. Also, the sensitivity to temperature changes of MRgHIFU systems can provide real-time thermal monitoring during treatments [15-18]. However, MRgHIFU needs large and expensive systems in comparison with USgHIFU, which is relatively cheap and portable [19]. In addition, considering the risk of skin burns in some MRgHIFU surgery cases [20-22], USgHIFU is more recommended in clinical tumor treating. When USgHIFU systems are equipped with specially designed transducers, it is possible to realize real-time visual temperature monitoring of the targeted soft tissue [23].

It is hard to realize the long-time thermostatic hyperthermia and keep the temperature constant with little fluctuation [24] because of lacking precisely adjustable heating and temperature measurement techniques [25]. In our study, a multi-elements ring ultrasound transducer array is used to generate spatially variable focus that provides accurate HIFU heating for tumor at any position in human thigh. Pulse signals are transmitted instead of continuous wave to ensure better focus effect since using continuous wave signals may increase the difficulty in accurate thermostatic control. Our proposed thermostatic hyperthermia model works on the heat conduction equation considering vascular heat perfusion combined with the preset equation of temperature and heat source. According to the change of temperature during the last heating period, the heat source for the consequent heating period is changed accordingly to realize the accurate control of constant temperature. Based on the known relationship between temperature and SOS, the measurement of the temperature in heating area is converted to read the value in SOS image which is updated accordingly. In our proposed model, the temperature measurement can share the same ultrasound transducer array for HIFU heating and B-mode ultrasound imaging, making the whole hyperthermia system concise and effective, which is of great guiding significance when this model is put into practical clinical application.

## II. MATHEMATICAL MODEL AND THEORY

### A. Acoustic Heating

Acoustic waves are the propagation form of sound, essentially the transmission of energy in the medium. When acoustic waves propagate, the classic case is to pass through a lossless fluid medium with homogeneous density, which satisfies the simple conservation of momentum, mass and energy [26]. But the reality is that the acoustic medium is heterogeneous, and its SOS and ambient density will vary. At the same time, the propagation of acoustic waves cannot ignore the loss of acoustic energy due to random thermal motion. In addition, acoustic waves passing through the compressible medium will cause dynamic fluctuations in variables such as density, velocity of acoustic particle, pressure, and temperature, etc. The vibration and friction between the acoustic waves and the medium will convert a part of the acoustic energy into thermal energy to generate acoustic absorption physically with dispersion [27].

Since we use a ring ultrasound transducer array in this study, acoustic heating is realized by a dynamic delayed emission of acoustic pulsed for each ultrasound transducer elements according to Eq. (1):

$$(x_i - o_{x0})^2 + (y_i - o_{y0})^2 = R^2,$$
$$l_i = \frac{\sqrt{(x_i - o_x)^2 + (y_i - o_y)^2}}{v_s},$$
$$t_0 = \frac{l_0}{v_s},$$
$$\Delta t_i = t_0 - \frac{l_i}{v_s}, \quad (1)$$

where $(x_i, y_i)$ is the coordinate of single element on ring ultrasound transducer array with the radius of R, which has 256 elements ($i$ ranges from 1 to 256). $(o_{x0}, o_{y0})$ is both the center of the ring ultrasound transducer array and the center of the human thigh model whereas $(o_x, o_y)$ is the focus position. In addition, $v_s$ stands for the SOS of tissue. We calculate the distance $l_i$ from each element to the focus and take the time $t_0$ calculated from the farthest distance $l_0$ as a reference, which can ensure that the delay value of each element $\Delta t_i$ is greater than or equal to 0 when transmitting signals. When we add the corresponding delay $\Delta t_i$ to each ultrasound transducer element, all the transmitted signals can reach the focus of the targeted position at the same time.

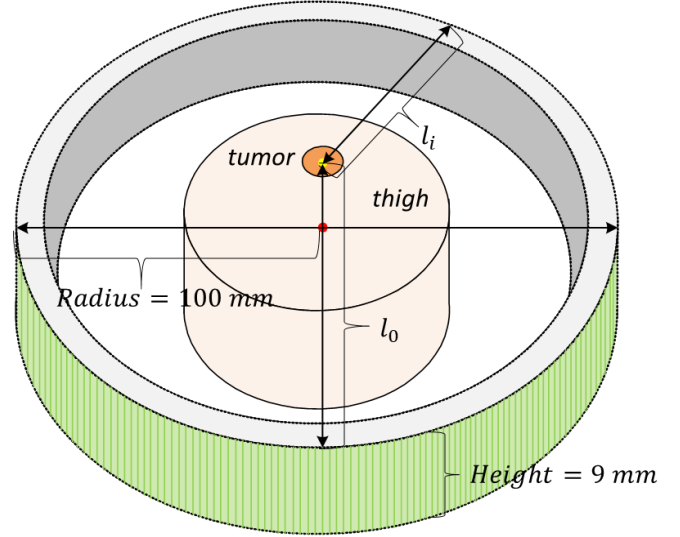

Fig.1. Schematic illustration of the ring transducer array. The green ring on the periphery stands for the ring ultrasound transducer array of 256 elements. The flesh-colored cylinder indicates human thigh whereas the orange area indicates the tumor. Furthermore, the red dot is the center of the thigh cross section and the yellow dot is the center of tumor.

The ring transducer array applied in this study is illustrated in Fig.1 which is fully consistent with the actual experimental equipment provide by Wayne State University, Detroit, MI, 48202, USA. In addition, both the simulation and the future experimental equipment are immersed in the water bath at a constant temperature at 30°C.

The heat conduction equation is an important partial differential equation in HIFU heating, which describes how the temperature in a region varies over time. Thermal diffusion during heating is the main consideration. In addition, Pennes' bioheat equation [28] accounts for many aspects, such as heat deposition due to ultrasound absorption and advective heat loss due to tissue perfusion (blood flowing through the tissue). Therefore, the heat conduction equation considering the factor of vascular heat perfusion is given by

$$\frac{dT}{dt} = \frac{k}{c\rho} \cdot \nabla^2 T - \frac{B}{c\rho}(T - T_a) + \frac{Q}{c\rho}, \quad (2)$$

where $k$ is thermal conductivity in $W/m \cdot °C$, $c$ is specific heat capacity in $J/kg \cdot °C$, $\rho$ is medium density in $kg/m^3$. $Q$ is the volume rate of heat deposition in $W/m^3$, which can be treated as the initial heat source for thermostatic hyperthermia. Besides, $B$ is the product of blood density ($kg/m^3$), blood specific heat capacity ($J/kg \cdot °C$) and blood perfusion rate ($s^{-1}$) whereas $T_a$ is the blood ambient temperature set to 37°C by default. $\nabla^2 T$ represents the Laplace operator ΔT for spatial variables given by:

$$\nabla^2 T = \frac{\partial^2 T}{\partial x^2} + \frac{\partial^2 T}{\partial y^2} + \frac{\partial^2 T}{\partial z^2}, \quad (3)$$

which is second derivative of temperature to three spatial axes.

We establish a preset relationship between the temperature change $\Delta T$ (°C) and the heat source change $\Delta Q$ ($W/m^3$) in Eq. (4), which is the key to thermostatic hyperthermia model.

$$\begin{aligned}\Delta Q &= -m \cdot \Delta T, \\ m &= m - \alpha 1 \cdot m, \\ m &= m + \alpha 2 \cdot m,\end{aligned} \quad (4)$$

where $m$ is the core operator that allows the thermostatic system to change the size of the heat source according to the temperature change. Besides, $\alpha$ is the control factor of $m$. When the temperature rises during the last heating period, $\Delta T$ is positive and $\Delta Q$ is negative correspondingly, so it is necessary to reduce the heat source in the consequent heating period. When the temperature drops, the process is similar. Using continuous heating by adjusting the size of heat source ensures that the temperature fluctuates up and down. To achieve the objective that the temperature fluctuation is within the desired range of the targeted temperature, it is important to fine-tune $m$. We first set a smaller upper and lower limit according to the targeted temperature range. If the temperature exceeds the preset upper limit, $m$ will decrease by $\alpha 1$ times of the previous-period $m$. The increase by $\alpha 2$ times of the previous-period $m$ occurs when the temperature exceeds the preset lower limit. The increasing thermal diffusion over time requires the control factor $\alpha$ to strengthen the control of $m$. Therefore, as long as the temperature exceeds the upper or lower limit, $\alpha 1$ and $\alpha 2$ will respectively add 0.01 to their own current values. Through combining the above equations, the model can effectively ensure that the temperature fluctuation is within a certain controllable range.

*B. Speed of Sound Imaging*

Speed of sound (SOS) imaging is a non-invasive imaging modality that can show the acoustic property change, such as the speed sound, in body tissue [29]. Previous studies have shown the sound speed in body tissue is sensitive to the temperature change [30]. In our study, the temperature control is necessary to the thermostatic hyperthermia system. Due to the available relationship between the temperature and the SOS (the temperature and the SOS vary in direct proportion) [31], the temperature monitoring in the medium is realized by the SOS imaging. There are basically three categories of SOS imaging methods. The first one is filtered back projection (FBP) reconstruction technique, which is also one of the main techniques in image reconstruction in clinical CT applications [32]. By employing the Radon transform, FBP technique can complete the image reconstruction in frequency domain, which avoid the direct solution of complex inverse problem consisting in finding the mapping function that matches the measured projection data and the sound speed distribution. In this way, the FBP reconstruction technique can provide fast computing speed with the reduce of computation complexity by using the

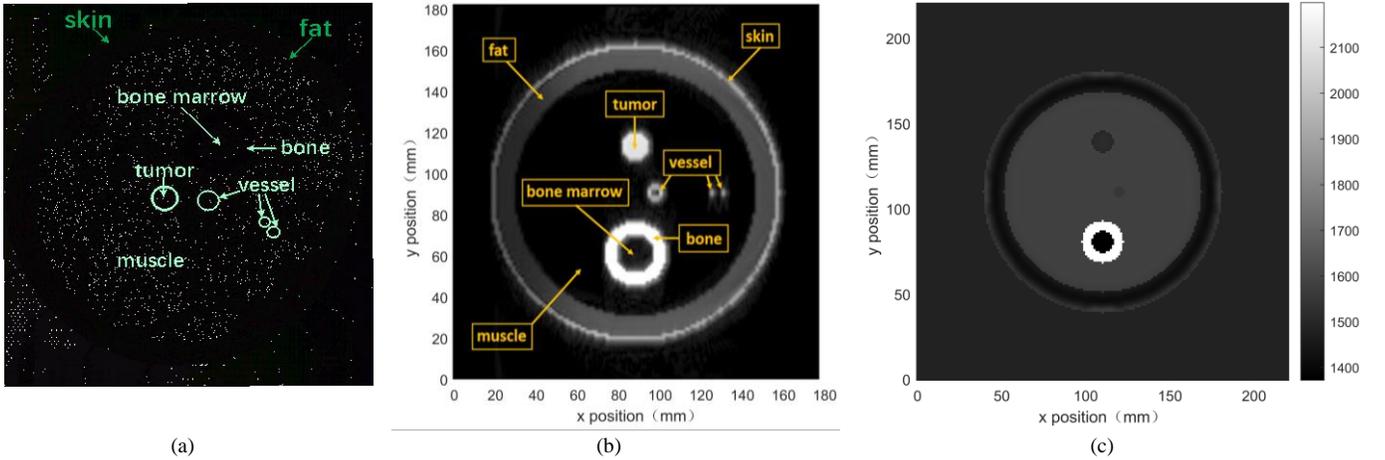

Fig.2. (a) Cross-sectional anatomy image of human thigh tissue. (b) B-mode ultrasound imaging of human thigh tissue. (c) Reconstruction image of SOS imaging.

Radon transform, which is the fundament of filtered back projection algorithm. High quality image can be produced when enough transmission data have been acquired, which is difficult to realize in practical application. Besides, FBP technique assumes that the propagation of acoustic wave travels along straight lines in body tissue [33]. The approximation causes the limitation of the precision of SOS reconstruction. The second one is bent-ray tracing method (BRTM) which is based on ray theory, which employs time of flight (TOF) of acoustic wave propagation paths to the sound speed distribution reconstruction [34]. In this method, the acoustic wave propagation paths can be estimated using the principles of geometrical optics and are not straight when traveling through different medium interface [35]. The precision of the TOF measurement is crucial to the accuracy of sound speed distribution estimation. In our application, the average length of acoustic wave propagation paths is around 120mm, the average SOS is about 1580m/s. Assuming that the sampling frequency of ultrasound transducer is 20MHz, the sound speed average error is more than 1m/s, which means that the deviation of temperature monitoring is not content with the temperature fluctuation limitation. The third one is called full-wave inversion (FWI), based on the acoustic wave propagation theory, which has many applications in medical imaging and geophysics [36-38]. When applying FWI method, the SOS image reconstruction is realized by estimating the SOS distribution using the acoustic wave equation and the measured data. The measured acoustic wave data can be acquired at the receiver ultrasound sensor elements and the estimated acoustic wave data can be acquired through pseudospectral k-space method. Given the gradient of a sum of squared norms between the measured data and the estimated data, a gradient descent-based optimization algorithm can be employed to update the sound speed estimate at every iteration [39]. Considering high order diffraction and scattering, FWI method can provide the better image quality in robustness, resolution and accuracy compared with previous two methods that have approximate hypothesis [40]. In our work, the temperature monitoring in the soft tissue is realized by the FWI method, which depends on the optimization iteration algorithm to achieve the SOS distribution reconstruction. At the beginning of iterations, the initial assumed SOS values of the medium plays an important role in calculation speed and accuracy of SOS image reconstruction. Through a correspondent B-mode ultrasound image, the boundary information and position information of human thigh tissues can be acquired. Combined with cross-sectional anatomy image of human thigh and acoustic properties of human body, the precise initial values of SOS can be employed as the priori information to estimate the actual SOS values, which is helpful to improve the calculation efficiency.

## III. EXPERIMENT AND ANALYSIS

### A. Simulation Tools

Our simulation mainly relies on K-Wave to build model and perform HIFU heating. K-Wave as an open source toolbox can provide users a MATLAB script interface to simulate the time-domain propagation of acoustic waves in one-dimensional, two-dimensional or three-dimensional space. The core calculation model of the toolbox is the pseudospectral k-space scheme, which is used to calculate the spatial gradients. In addition to explaining linear and nonlinear acoustic propagation, K-Wave toolbox also has various functions including the calculating the distribution of heterogeneous material parameters and power law acoustic absorption [41-42].

In addition, we also use a Verasonics Vantage system (Vantage 64, Verasonics Inc., Kirkland, WA, USA) controlled by MATLAB in computer to perform B-mode ultrasound imaging on human thigh. The Verasonics Vantage system is a highly programmable platform that consists of the Vantage software, the Vantage hardware simulator and a computer. Users can edit and run the Verasonics scripts on the Matlab interface to perform beamforming and postprocessing of signals through the Vantage software. The Vantage hardware is responsible for collecting data sequences and calculating RF backscatter data with the support of some circuit boards and acquisition Boards [43].

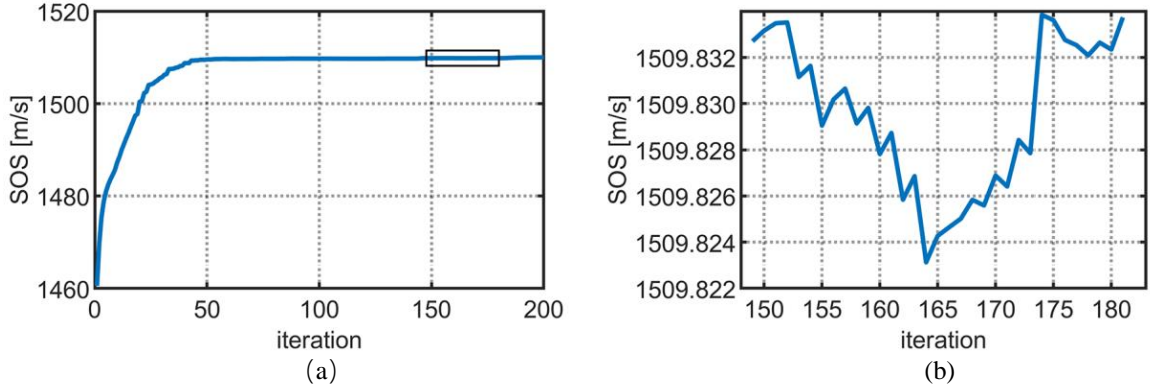

Fig.3.(a) The sound speed iteration of tumor region. (b) The average sound speed fluctuation of tumor region in the black rectangular box of Fig.3.(a).

### B. Experiment Setup

Our simulation is performed in a three-dimensional space supported by k-Wave. 3D simulation space is defined as $N_x \times N_y \times N_z$ ($490 \times 490 \times 40$) in grid size, in which the spacing of each grid point is about 0.449 mm. The actual physical size of the simulation model is $220\ mm \times 220\ mm \times 18\ mm$. By observing Fig.2(a), we can find that the internal structures of the human thigh including: skin, subcutaneous fat, muscle, thigh bone, bone marrow, tumor and blood vessels. All the tissues are defined respectively with reference to the properties such as size, SOS, density, attenuation coefficient, thermal conductivity and heat capacity in Table 1 [44]. In addition, the blood perfusion rate is set to $0.01 s^{-1}$. In our study, we assume the human thigh is placed in the ring ultrasound transducer array with 256 elements (the radius is 100 mm and the height is 9 mm). The whole thigh is immersed in water at a constant temperature at 30°C as shown in Fig.1. All the ultrasound elements transmit pulse signals of five cycles.

To improve the universality of the simulation, we set three thigh models, including 55.00mm, 60.00mm and 65.00mm of muscle radius. Furthermore, we change the positions of tumor according to its growth characteristic that tumors may grow close to blood vessels because they rely on vessels to provide a lot of oxygen and nutrients [45]. The model of human thigh referring to Fig.2(a) is given in Fig.4, which display the density map for different human thigh models.

*C. B-mode Ultrasound Imaging and Speed of Sound Imaging*

Fig.2(a) as a cross-sectional anatomy image provides a preliminary reference [46] for us to understand the human thigh structure. In order to build a simulation model with more details, the inner structure information of human thigh needs to be obtained by B-mode ultrasound imaging, especially around the tumor. Therefore, we make a B-mode ultrasound image of human thigh tissue through Verasonics Vantage system as shown in Fig.2(b), which helps us to find the focus position for HIFU heating and provides us with the priori information of the internal medium in SOS imaging. In the process of SOS image reconstruction, we first split up the human thigh model into seven parts including skin, fat, muscle, bone, bone marrow, tumor and vessels with the boundary information from the B-mode ultrasound image. The SOS in each part is unrelated with other parts and the initial SOS in each part refers to the acoustic properties of the corresponding human tissue. Furthermore, we set the SOS in bone, bone marrow and vessels to the standard values (SOS at room temperature) and keep the SOS in these regions constant. The reasons for the constant SOS in the above regions are as follows. Firstly, since the regions of bone and bone marrow are outside the effective heating range, HIFU heating almost has little influence in the two regions. Secondly, considering the size of vessel regions, the SOS change in the small region hardly contributes to the accuracy of the whole SOS iteration. Meanwhile, the SOS iteration in the small region is sensitive to noise which may lead to the wrong iteration direction of SOS, so it is necessary that SOS in vessel regions remains constant. These assumptions help us to improve the accuracy in estimating the SOS values and the reconstruction image of SOS is shown in Fig.2(c). Since the accurate SOS in tumor region is essential to the thermostatic hyperthermia, the SOS in tumor region is averaged to improve the robustness and accuracy of SOS estimation. During the SOS imaging iteration, the averaged SOS value in tumor region is illustrated in Fig.3. From Fig.3(b), which is a zooming curve of the black rectangle region in Fig.3(a), we can discern that the SOS fluctuation in the tumor region can be limited to less than 0.2m/s which in return can yield the temperature monitoring accuracy to 0.2℃.

*D. Long-time Thermostatic Hyperthermia*

Through HIFU focusing operation, the ultrasound beams with high intensity is focused in the center of the tumor to generate acoustic pressure field that demonstrates a good focusing effect. Long-term thermostatic hyperthermia can be divided into two heating stages. The first heating stage aims to quickly heat the tissue to the target temperature by continuously heating at full power until the temperature of the tumor center is close to the target temperature. The second heating stage is thermostatic control realized according to the heat conduction equation (as shown in Eq. (3)) and the equation of thermostatic hyperthermia model (as shown in Eq. (2)).

TABLE I
PROPERTIES OF MEDIUM

| Medium | Size ($mm$) | SOS ($m/s$) | Density ($kg/m^3$) | Attenuation Coefficient ($Np/Mhz \cdot m$) | Thermal Conductivity ($W/m \cdot ℃$) | Specific Heat Capacity ($J/kg \cdot ℃$) |
|---|---|---|---|---|---|---|
| Water | 220×220×18 (Volume) | 1482 | 1000 | 0.025 | 0.60 | 4178 |
| Skin | 1.70 (Thickness) | 1595 | 1109 | 21.158 | 0.37 | 3391 |
| Fat | 10.00 (Radius) | 1430 | 911 | 4.358 | 0.21 | 2348 |
| Muscle | 55.00/60.00/65.00 (Radius) | 1580 | 1090 | 7.109 | 0.49 | 3421 |
| Bone | 5.80 (Thickness) | 2198 | 1178 | 47.000 | 0.32 | 1313 |
| Bone Marrow | 6.70 (Radius) | 1372 | 980 | 4.358 | 0.20 | 2065 |
| Tumor | 6.50 (Radius) | 1450 | 1050 | 5.650 | 0.51 | 3540 |
| Big Vessel | 3.40 (Radius) | 1550 | 1060 | 2.368 | 0.52 | 3617 |
| Small Vessel | 1.15 (Radius) | 1550 | 1060 | 2.368 | 0.52 | 3617 |

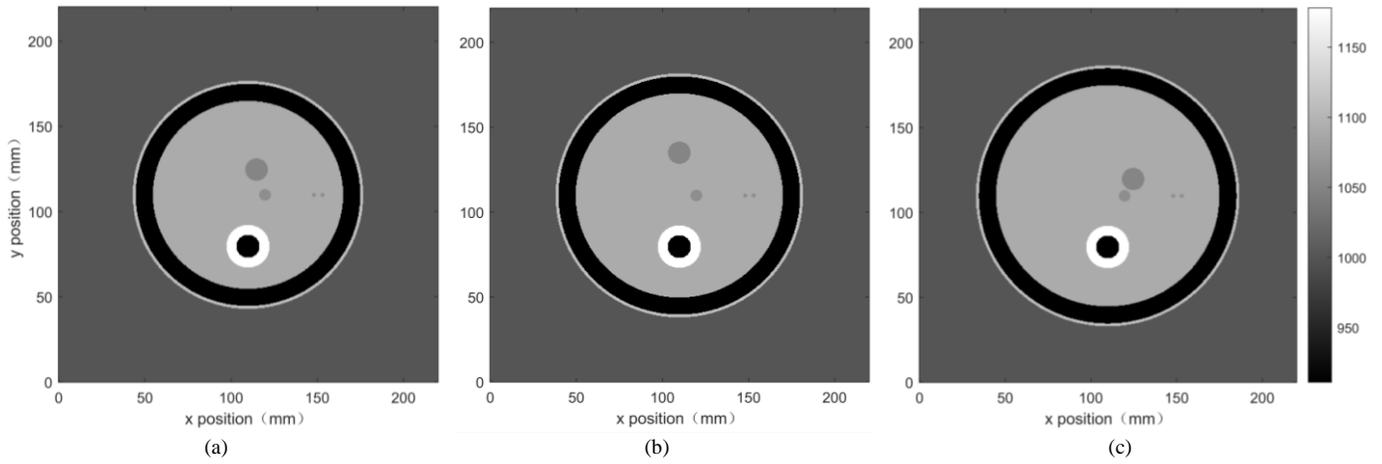

Fig.4. Density map of human thigh model. (a) Model A: radius of muscle is 55 mm. (b) Model B: radius of muscle is 60 mm. (c) Model C: radius of muscle is 65 mm.

The second heating stage is realized by dynamically changing the heating power according to Eq. (4) within each 60 seconds period. Operator $m$ is set to multiple initial values suitable for different $\Delta T$ based on the relation that $m$ and $\Delta T$ are inversely proportional (the product of $m$ and $\Delta T$ ranges from $8 \times 10^4 W/m^3$ to $2 \times 10^5 W/m^3$, which is also 0.005 to 0.01 times the amplitude of the previous-period heat source). According to Eq. (4), when the temperature of tumor center exceeds the upper and lower limit of the target temperature, the operator $m$ will self-adjust by $\alpha$ times of the previous-period $m$. It will self-decrease when temperature rises and exceeds the limit, and it will self-increase to prevent the temperature cooling down when temperature drops. The initial values of $\alpha 1$ and $\alpha 2$ in our study are both set to 0.01. Since the relationship between the temperature and the SOS in the different human tissue is available (the temperature and the SOS vary in direct proportion), the temperature detection in the last heating period is converted from the SOS value which is updated accordingly. We set two targeted temperatures 42°C and 43°C respectively for each model and the temperature control curve of long-term thermostatic hyperthermia is illustrated in Fig.5. It can be observed that the temperature can be kept within $\pm 0.2°C$ range of the targeted temperature with slight fluctuations for 5 hours. Furthermore, we also concerned about the temperature distribution in three-dimensional perspective during the whole heating process. We select model C in Fig.4(c) with the targeted temperature of 43°C as an observation example. As shown in the Fig.6, the heating is mainly concentrated in the center of the tumor and there is almost no obvious heat deposition in other area, which proves that the HIFU heating can achieve precise focusing on the tumor. Furthermore, the temperature in other areas quickly drops to the body temperature after the first stage of heating, which proves the effectiveness of our proposed thermostatic hyperthermia model.

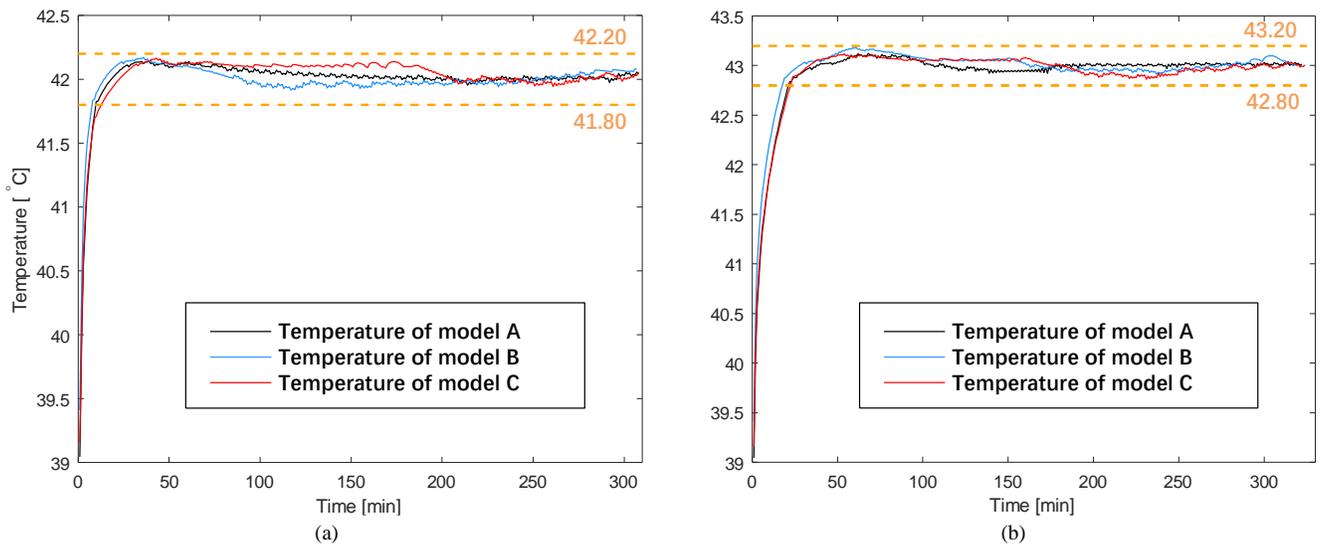

Fig.5. Temperature control curve at tumor center for about 5 hours' HIFU heating. (a) Three models using 42°C as the target temperature. (b) Three models using 43°C as the target temperature.

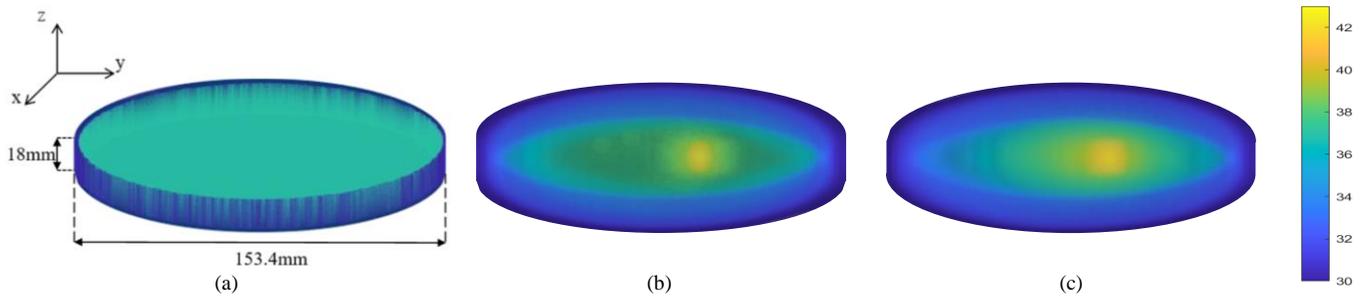

Fig.6. 3D temperature distribution at 3 different time with the same size. (a) Initial temperature distribution. (b) Temperature distribution at the end of the first heating stage. (c) Temperature distribution at steady state.

IV. CONCLUSION

This paper proposes a thermostatic hyperthermia model for tumor therapy with non-invasive temperature measurement through SOS imaging. We use the ring ultrasound transducer array to achieve the precise and variable focus position with good focusing effect at the tumor center. Also, the results of temperature control process prove that we have solved the critical problem of long-time thermostatic hyperthermia with the temperature being kept constant. Furthermore, the temperature monitoring and the thermostatic hyperthermia sharing the same system are designed to guide the production of corresponding medical equipment, which realize the integration of functions and reduce the difficulty of clinical implementation. In the future, when our study is combined with actual experiments, it is possible to extend the tumor treatment to other parts of human body, thereby providing promising clinical application value.